\documentclass{emulateapj}

\begin{document}
\slugcomment{Accepted to ApJ}

\title{ MMT Hypervelocity Star Survey }

\author{Warren R.\ Brown,
	Margaret J.\ Geller,
	Scott J.\ Kenyon}

\affil{Smithsonian Astrophysical Observatory, 60 Garden St, Cambridge, MA 02138}
\email{wbrown@cfa.harvard.edu} 
 % mgeller@cfa.harvard.edu, skenyon@cfa.harvard.edu}

\shorttitle{ MMT Hypervelocity Star Survey }
\shortauthors{Brown, Geller, \& Kenyon}

\begin{abstract}

	We describe a new survey for unbound hypervelocity stars (HVSs), stars
traveling with such extreme velocities that dynamical ejection from a massive black
hole is their most likely origin.
	We investigate the possible contribution of unbound runaway stars, and show
that the physical properties of binaries constrain low mass runaways to bound
velocities.
	We measure radial velocities for HVS candidates with the colors of
early A-type and late B-type stars.
	We report the discovery of 6 unbound HVSs with velocities and distances
exceeding the conservative escape velocity estimate of Kenyon and collaborators.  
We additionally report 4 possibly unbound HVSs with velocities and distances
exceeding the lower escape velocity estimate of Xue and collaborators.  These
discoveries increase the number of unbound HVSs by 60\% - 100\%.
	Other survey objects include 19 newly identified $z\sim2.4$ quasars.
	One of the HVSs may be a horizontal branch star, consistent with the number
of evolved HVSs predicted by Galactic center ejection models.  Finding more evolved
HVSs will one day allow a probe of the low-mass regime of HVSs and will constrain
the mass function of stars in the Galactic center.

\end{abstract}

\keywords{
        Galaxy: halo ---
        Galaxy: center ---
        Galaxy: stellar content ---
        Galaxy: kinematics and dynamics ---
        stars: early-type
}

\section{INTRODUCTION}

	Three-body interactions with a massive black hole (MBH) will inevitably
unbind stars from the Galaxy \citep{hills88, yu03}.  In 2005 we reported the
discovery of the first HVS:  a 3 M$_{\sun}$ main sequence star traveling with a
Galactic rest frame velocity of at least $+700\pm12$ km s$^{-1}$, more than twice
the Milky Way's escape velocity at the star's distance of 110 kpc \citep{brown05}.  
This star cannot be explained by normal stellar interactions: the maximum ejection
velocity from binary disruption mechanisms \citep{blaauw61, poveda67} is limited to
$\lesssim$300 km s$^{-1}$ for 3 M$_{\sun}$ stars \citep{leonard91, leonard93,
tauris98, portegies00, davies02, gualandris05}.  Although runaways may reach unbound
velocities for very massive stars, like the hyper-runaway HD 271791 \citep{heber08,
przybilla08c}, runaway ejection velocities are constrained by the properties of
binary stars. A massive and compact object is needed to accelerate low mass stars to
unbound velocities.

	There is overwhelming evidence for a $\sim4\times10^6$ M$_{\sun}$ MBH in the
dense stellar environment of the Galactic center \citep{schodel03, ghez08}.  
\citet{hills88} coined the term HVS to describe a star ejected by the MBH.  The
observational signature of a HVS is its unbound velocity.  Although not all unbound
stars are necessarily HVSs -- fast-moving pulsars, for example, are explained by
supernova kicks \citep[e.g.][]{arzoumanian02} -- unbound low mass main sequence
stars are most plausibly explained as HVSs.

	Here we introduce a new HVS survey using the MMT to target HVS candidates
with masses down to $\sim$2 M$_{\sun}$.  Discovering lower-mass HVSs should provide
constraints on the stellar mass function of HVSs \citep{brown06, kollmeier07}; the
velocity distribution of low- versus high-mass HVSs may discriminate between a
single MBH or binary MBH origin \citep{sesana07b, kenyon08}.  Our survey strategy
targets stars fainter and redder than the original HVS survey.  This strategy is
successful:  we report the discovery of 6 unbound HVSs and 4 possibly unbound HVSs.

\subsection{Recent Observations of HVSs}

	Observers have identified a remarkable number of HVSs in the past 3 years.  
Following the discovery of the first HVS \citep{brown05}, \citet{hirsch05} reported
a helium-rich subluminous O star leaving the Galaxy with a rest-frame velocity of at
least $+717$ km s$^{-1}$.  \citet{edelmann05} reported an 9 M$_{\sun}$ main sequence
B star with a Galactic rest frame velocity of at least $+548$ km s$^{-1}$, possibly
ejected from the Large Magellanic Cloud.  \citet{brown06, brown06b, brown07a,
brown07b} reported 7 B-type HVSs discovered in a targeted HVS survey, along with
evidence for an equal number of bound HVSs ejected by the same mechanism.

	High-dispersion spectroscopy has shed new light on the nature of the HVSs.
\citet{przybilla08b} have recently shown that HVS7 is a chemically peculiar B main
sequence star, with an abundance pattern unusual even for the class of peculiar B
stars.  The star HVS3 (HE 0437-5439), the unbound HVS very near the LMC on the sky,
has received the most attention.  HVS3 is a 9 M$_{\odot}$ B star of half-solar
abundance, a good match to the abundance of the LMC \citep{bonanos08, przybilla08}.  
Stellar abundance may not be conclusive evidence of origin, however.  A- and B-type
stars exhibit 0.5 - 1 dex scatter in elemental abundances within a single cluster,
due to gravitational settling and radiative levitation in the atmospheres of the
stars \citep{varenne99, monier05, fossati07, gebran08a, gebran08b}.

	An LMC origin requires that HVS3 was ejected from the galaxy at $\sim$1000
km s$^{-1}$ \citep{przybilla08}, a velocity that can possibly come from three-body
interactions with an intermediate mass black hole in a massive star cluster
\citep{gualandris07, gvaramadze08}.  \citet{perets08b} shows that the ejection rate
of 9 M$_{\sun}$ stars, however, is four orders of magnitude too small for this
explanation to be plausible.  The alternative explanation is that HVS3 is a blue
straggler, ejected by the Milky Way's MBH.  Theorists argue that a single MBH or a
binary MBH can eject a compact binary star as a HVS \citep{lu07, perets08b}; the
subsequent evolution of such a compact binary can result in mass-transfer and/or a
merger that can possibly explain HVS3 \citep{perets08b}.  Proper motion
measurements, underway now, will determine HVS3's origin.

	Other recent HVS work highlights the link between stellar rotation and the
origin of HVSs.  Main sequence B stars have fast mean $v\sin{i}\sim150$ km s$^{-1}$
\citep[e.g.][]{abt02, huang06a}.  Hot blue horizontal branch (BHB) stars have slow
mean $v\sin{i}<10$ km s$^{-1}$ (because they have just evolved off the red giant
branch \citep[e.g.][]{behr03, behr03b}).  Interestingly, \citet{hansen07} predicts
that main sequence HVSs ejected by the Hills mechanism should be slow rotators,
because stars in compact binaries have systematically lower $v\sin{i}$ due to tidal
synchronization.  \citet{lockmann08}, on the other hand, predict that HVSs should be
fast rotators, at least for single stars spun up and ejected by a binary black hole.  
To date HVS1, HVS3, HVS7, and HVS8 have observed $v\sin{i}$ of $\sim$190, $55\pm2$,
$55\pm2$, and $260\pm70$ km s$^{-1}$, respectively \citep{heber08b, przybilla08,
przybilla08b, lopezmorales08}.  As discussed by \citet{perets07c}, we clearly
require a larger sample of HVSs to measure the distribution of HVS rotations and
discriminate HVS ejection models.

	In \S 2 we describe our new HVS survey strategy and summarize our
observations.  In \S 3 we discuss the Galactic escape velocity, our definition of a
HVS, and the possible contribution of hyper-runaways to the population of unbound
stars.  In \S 4 we present the new unbound HVSs.  In \S 5 we discuss a possible BHB
star among the HVSs.  We conclude in \S 6.

\begin{figure}		% FIGURE 1: COLOR-COLOR PLOT
 \plotone{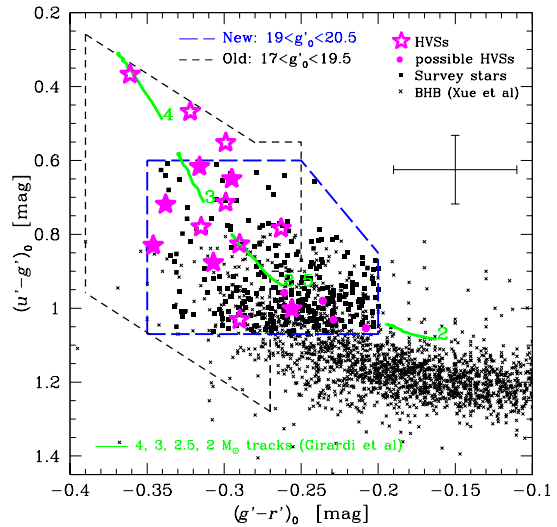}
 \caption{ \label{fig:ugr}
	Color-color diagram illustrating the target selection for our new HVS survey
({\it long dashed line}) and our old HVS survey ({\it short dashed line}).  The six
new HVSs ({\it solid stars}) and four possible HVSs ({\it solid dots}) scatter
around the \citet{girardi04} stellar evolution tracks for 2 - 4 M$_{\sun}$ main
sequence stars ({\it solid lines}).  Average color uncertainties are indicated by
the errorbar on the upper right.  Previous HVS discoveries ({\it open stars}) and
the \citet{xue08} BHB sample ({\it x's}) are also marked.}
 \end{figure}

\section{DATA}

\subsection{New Target Selection}

	HVSs are rare: our discoveries imply there are $96\pm20$ HVSs of mass 3-4
M$_{\odot}$ within 100 kpc \citep{brown07b}.  Thus to find new HVSs we must target
luminous objects over a very large volume.
	Luminous O- and B-type HVSs, even if they lived long enough to be observed,
would be lost behind a large foreground of hot white dwarfs with identical colors,
$(u'-g')_0 \lesssim 0.4$.  We design our new HVS survey to target early A-type
stars.

	Known HVSs are located at distances 50 - 100 kpc, corresponding to $19.5 <
g'_0 < 21$ for early A-type stars.  At such magnitudes, fewer than $\sim$100 A-type
stars have published radial velocities \citep{sirko04a, clewley05, xue08}.  Halo BHB
and blue straggler stars are a major contaminant at faint magnitudes, but
fortunately the density of the stellar halo falls off very steeply.  
\citet{kenyon08} calculate the density profile of unbound HVSs is approximately
$\rho \propto r^{-2}$ \citep[see also][]{kollmeier07}; the density profile of the
stellar halo is closer to $\rho \propto r^{-3}$ \citep[e.g.][]{juric08}. Thus we
maximize the contrast of HVSs with respect to indigenous halo stars by restricting
ourselves to the faintest A-stars.

	We select candidate HVSs in the magnitude range $19.0<g'_0<20.5$.  The faint
limit is set by Sloan Digital Sky Survey (SDSS) photometric errors, which approach
$\pm0.15$ in $(u'-g')_0$ at $g'=20.5$.  The bright limit is set to provide
continuity with our original HVS survey, although we impose a more stringent bright
limit on the reddest stars $g'_0 > 18.6 + 10[(g'-r')_0+0.3] +
[(u'-g')_0-(1.2(g'-r')_0+1.25)]$.  In other words, $g'_0>19.5$ at $(g'-r')_0=-0.2$.  
We use a combination of $(u'-g')_0$ and $(g'-r')_0$ colors to select stars with
probable high surface gravity, and thus maximize the chance of finding main
sequence HVSs.

	Figure \ref{fig:ugr} illustrates the color selection.  The original HVS
survey (dashed line) was designed to avoid the locus of BHB/A-type stars; our new
HVS survey (long dashed line) follows the \citet{girardi02, girardi04} main sequence
tracks for solar abundance stars (solid lines) down to $\sim$2 M$_{\odot}$ A-stars.  
Known HVSs scatter uniformly around the main sequence tracks.  Known BHB stars
\citep[][]{xue08}, on the other hand, are systematically redder in $(u'-g')_0$.

	We select stars with $0.6 < (u'-g')_0 < 1.07$ to avoid the majority of known
halo BHB stars (Figure \ref{fig:ugr}).  We select stars with $-0.35 < (g'-r')_0 <
-0.20$ to include 3 M$_{\odot}$ stars that can viably travel 150 kpc (=20.5 mag) in
their main sequence lifetimes.  Finally, we impose $-0.5<(r'-i')_0 <0$ and
$(g'-r')_0 < 0.2(u'-g')_0 -0.38$ to exclude non-stellar objects, such as quasars.  
Notably, this color selection includes the first HVS, that was not formally a part
of the original HVS survey.

	Applying this color-magnitude selection to the SDSS DR6 photometric catalog
results in 528 HVS candidates spread over 7300 deg$^2$.  We have excluded the small
region of the SDSS between $b<-l/5 + 50\arcdeg$ and $b>l/5-50\arcdeg$ to avoid
excessive contamination from Galactic bulge stars.  Of the 528 HVS candidates, 59
were previously observed as part of our original HVS survey \citep{brown07b}, and 21
are in the \citet{sirko04a} BHB catalog.  Thus we need spectra for 448 HVS
candidates.

\subsection{Spectroscopic Observations}

	Spectroscopic observations were obtained at the 6.5m MMT telescope with the
Blue Channel spectrograph on the nights of 2008 February 6-10 and 2008 May 7-11.  
We operated the spectrograph with the 832 line mm$^{-1}$ grating in second order and
a 1.25$\arcsec$ slit.  These settings provide wavelength coverage 3650 \AA\ to 4500
\AA\ and a spectral resolution of 1.2 \AA.  All observations were obtained at the
parallactic angle.

	Our goal was to obtain modest signal-to-noise ($S/N$) observations adequate
for determining radial velocity.  We typically obtained $S/N=5$ in the continuum at
4000 \AA\ in a 10 minute integration on a $g'=20$ star.  We obtained spectra
for 233 HVS candidates, and processed the data in real-time to allow additional 
observations of interesting candidates.
	We extracted the spectra using IRAF\footnote{IRAF is distributed by the
National Optical Astronomy Observatories, which are operated by the Association of
Universities for Research in Astronomy, Inc., under cooperative agreement with the
National Science Foundation.}
	in the standard way and measured radial velocities using the
cross-correlation package RVSAO \citep{kurtz98}.  The average radial velocity
uncertainty of the $S/N=5$ spectra is $\pm20$ km s$^{-1}$.

\subsection{HVS Sample}

	We now have spectroscopic identifications and radial velocities for 313
(59\%) of the 528 HVS candidates.  19 of the objects are newly identified $z\sim2.4$
quasars, and 9 of the objects are DA white dwarfs.  We present the quasars and white
dwarfs in Appendix A.  The remaining 285 objects have the spectra of early A- and
late B-type stars.  In addition, we observed the final 40 objects remaining in the
original HVS survey \citep{brown07b}.

	Because our new and original HVS surveys cover contiguous regions of
color-magnitude space over the same region of sky, we consider the results of the
combined surveys in this paper.  We exclude stars with $g'_0<17$ that are possibly
associated with the inner halo; the inner halo has distinctly different kinematics
from the outer halo \citep{carollo07, morrison08}.  We also exclude all white
dwarfs, quasars, and B supergiants.  Our combined HVS survey contains 759
non-kinematically-selected stars $17<g'_0<20.5$.

\begin{figure}		% FIGURE 2: VEL HISTOGRAM
 \plotone{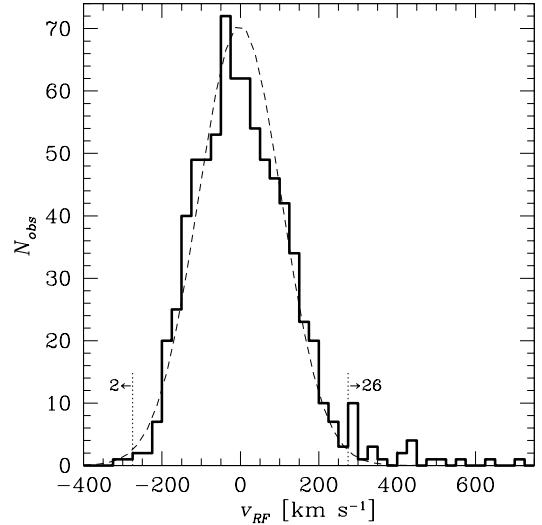}
 \caption{ \label{fig:hist}
	Minimum Galactic rest-frame velocity $v_{rf}$ distribution for the 759 stars
in the combined HVS survey.  The best-fit Gaussian ({\it dashed line}) has
dispersion $106\pm5$ km s$^{-1}$, excluding the 26 stars with $v_{rf}>+275$ km
s$^{-1}$.  The asymmetry of positive velocity outliers is significant at the
$\sim$$4\sigma$ level and shows that the $+300$ km s$^{-1}$ stars are short-lived;
we observe only 2 star falling back onto the Galaxy around $-300$ km s$^{-1}$. }
 \end{figure}

\subsection{Radial Velocity Distribution}

	Figure \ref{fig:hist} plots the distribution of line-of-sight velocities,
corrected to the Galactic rest-frame \citep[see][]{brown06b}, for the combined
sample of 759 stars.  
	The 731 survey stars with $|v_{rf}|<275$ km s$^{-1}$ have a $-1\pm4$ km
s$^{-1}$ mean and a $106\pm5$ km s$^{-1}$ dispersion, consistent with a normal
stellar halo population.  Notably, there are 28 stars in the tails of the
distribution with $|v_{rf}|>275$ km s$^{-1}$.

	We observe 26 stars with $v_{rf}>275$ km s$^{-1}$ and only 2 stars with
$v_{rf}<-275$ km s$^{-1}$.  The escape velocity of the Milky Way at 50 kpc is
$\sim360$ km s$^{-1}$ \citep{kenyon08}, thus the 12 stars with $v_{rf}>400$ km
s$^{-1}$ are unbound.
	Ignoring the 12 unbound stars, there is less than a $10^{-5}$ probability of
randomly drawing 14 stars with $275<v_{rf}<400$ km s$^{-1}$ from the tail
of a Gaussian distribution with the observed parameters.  Thus the excess of
positive velocity outliers $275<v_{rf}<400$ km s$^{-1}$ appears significant at the
4-$\sigma$ level.

	The positive velocity outliers demonstrate a population of short-lived HVSs
ejected onto bound trajectories \citep{brown07a, brown07b}.  HVS ejection mechanisms
naturally produce a broad spectrum of ejection velocities \citep[e.g.][]{
sesana07b}.  Simulations of HVS ejections from the Hills mechanism suggest there
should be comparable numbers of unbound and bound HVSs with $v_{rf}>+275$ km
s$^{-1}$ in our survey volume \citep{bromley06}.  As shown below, we find 14 unbound
HVSs and 12 possibly bound HVSs with $v_{rf}>+275$ km s$^{-1}$, in good agreement
with model predictions.

\section{UNBOUND STARS}

\subsection{Defining Hypervelocity Stars}

	Following \citet{hills88}, we define HVSs by 1) their MBH origin and 2)  
their unbound velocities.
	An HVS ejected from the Milky Way travels on a nearly radial trajectory; the
expected proper motion for a 50 kpc distant HVS is a few tenths of a milliarcsecond
per year \citep[e.g.][]{gnedin05}.  Thus radial velocity directly measures most of a
HVS's space motion.  Deciding whether a HVS is unbound, however, requires knowledge
of the star's distance.

	We estimate distances to the HVSs assuming they are main sequence stars.  
This assumption is motivated by the significant absence of stars falling back onto
the Galaxy around $-300$ km s$^{-1}$ (Figure \ref{fig:hist}).  Bound HVSs must have
main sequence lifetimes less than $\sim$1 Gyr, otherwise we would see them falling
back onto the Galaxy \citep{brown07b, kollmeier07, yu07}.  Given the color-selection
of our survey (Figure \ref{fig:ugr}), the A- and B-type HVSs must be 2-4 M$_{\odot}$
main sequence stars at Galactocentric distances $>40$ kpc (Figure \ref{fig:colmags}).

\begin{figure}		% FIGURE 3: TRAVEL TIME HISTORY
 \plotone{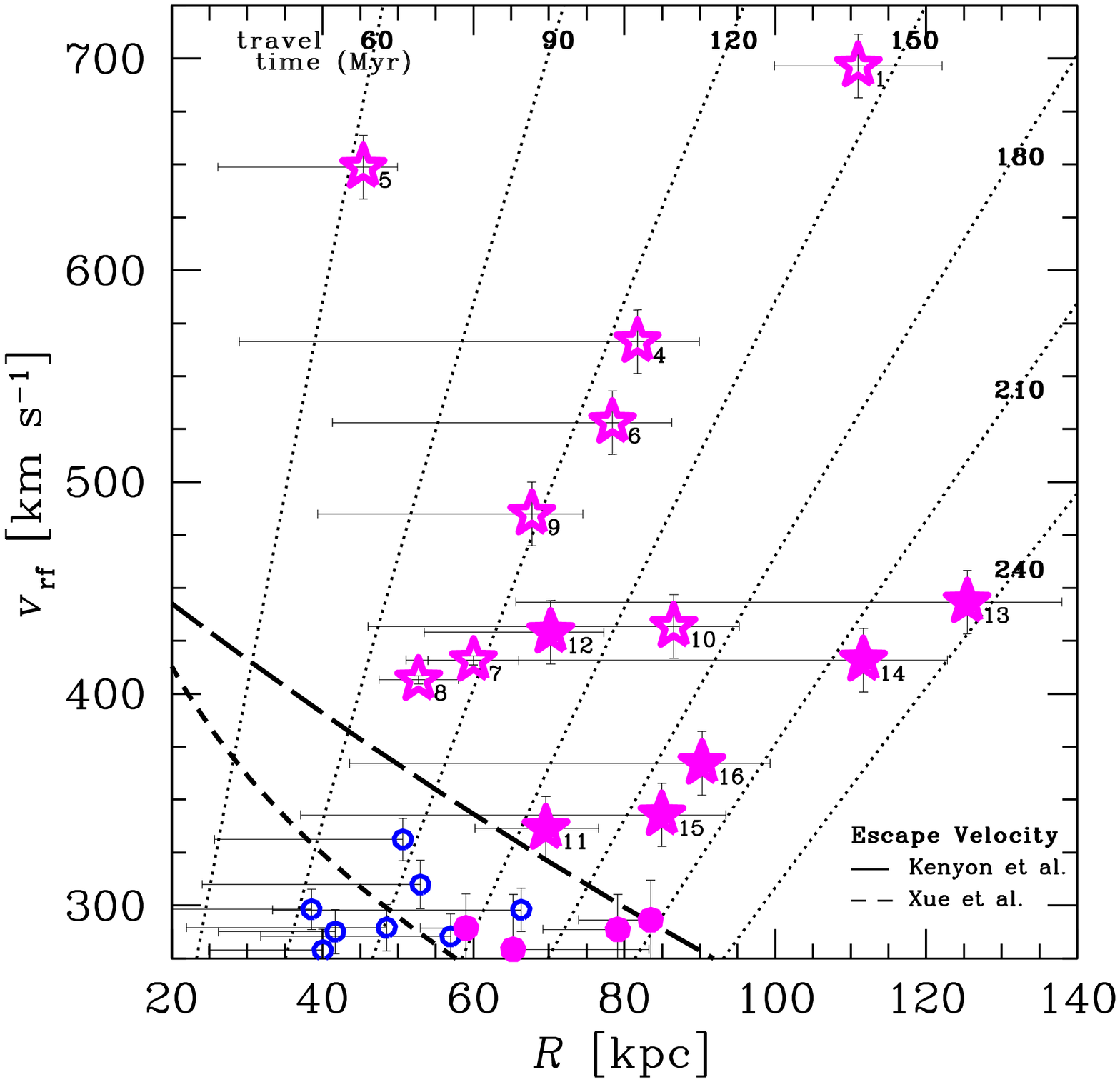}
 \caption{ \label{fig:travel}
	Minimum rest-frame velocity vs.\ Galactocentric distance $R$ for the 26
stars with $v_{rf}>+275$ km s$^{-1}$.  The six new HVSs ({\it solid stars}) have
velocities and main sequence star distances exceeding the \citet{kenyon08} escape
velocity model ({\it long dashed line}).  Four possible HVSs ({\it solid dots}) have
velocities, main sequence star distances, and BHB distances exceeding the
\citet{xue08} escape velocity model ({\it dashed line}).  Previous HVS discoveries
({\it open stars}) and possibly bound HVSs ({\it open circles}) are also indicated.  
Errorbars show the span of physically plausible distance if the HVSs were BHB stars.  
Isochrones of travel time from the Galactic center ({\it dotted lines})  are
calculated using the potential model of \citet{kenyon08}, assuming the observed
minimum rest frame velocity $v_{rf}$ is the full space motion of the stars.  }
 \end{figure}

	Unfortunately, the Galactic potential is poorly constrained at distances
$>40$ kpc.  We consider two Galactic potential models here.  \citet{kenyon08}
discuss a spherically symmetric potential that, for the first time, fits the Milky
Way mass distribution from 5 pc to 10$^5$ pc.  Because the form of the potential
does not yield a true escape velocity, \citet{kenyon08} define unbound stars as
having $v_{rf}>200$ km s$^{-1}$ at $R=150$ kpc.  This conservative definition yields
a Galactic escape velocity of 360 km s$^{-1}$ at 50 kpc and 260 km s$^{-1}$ at 100
kpc (Figure \ref{fig:travel}).  \citet{xue08}, on the other hand, fit a halo
potential model to the velocity dispersion of 2466 BHB stars located 5 kpc $<R<60$
kpc.  The escape velocity resulting from the \citet{xue08} model is 290 km s$^{-1}$
at 50 kpc and 190 km s$^{-1}$ at 100 kpc (Figure \ref{fig:travel}).

\subsection{Hyper-Runaways}

	Not all unbound stars are HVSs.  The star HD 271791 is the first example of
an unbound ``hyper-runaway'' that was ejected from the outer disk, in the direction of
Galactic rotation, when its former 55 M$_{\sun}$ binary companion exploded as a
supernova \citep{heber08, przybilla08c}.  Objects ejected in this manner are
traditionally called runaways \citep{blaauw61}.  The term runaway also includes
stars dynamically ejected from binary-binary encounters \citep{poveda67}.

	We investigate here the possible contribution of runaways to the population
of HVSs.  First, we consider the properties of binaries required to produce unbound
hyper-runaways. Then, we consider the production rate of stars massive enough to
produce these hyper-runaways.

\subsubsection{Binary Star Properties}

	Both the supernova and binary-binary ejection mechanisms share a common
velocity constraint:  the physical properties of binary stars.
	Theoretically, the maximum ejection velocity from disrupting a binary (e.g.,
by a supernova) is the binary orbital velocity.
	The maximum ejection velocity from dynamical binary-binary encounters is the
escape velocity of the most massive star, which for stars on the upper main sequence 
is $v_{\rm esc} \simeq 700 (M/{\rm M}_{\sun})^{0.15}$ km s$^{-1}$ \citep{leonard91}.
	This theoretical maximum ejection velocity is not realizable, however,
because compact binaries that are too tight will merge instead of producing a
runaway.

	To avoid merging, a compact binary must avoid losing too much energy from
Roche Lobe overflow and from tidal dissipation.
	Stars with separations less than 2.5 $R_{\rm star}$ overfill their Roche
lobes and quickly merge \citep[e.g.][]{vanbeveren98}.  During close binary
encounters, tidal dissipation can lead to mergers of compact binaries \citep{lee86,
mcmillan87, leonard88}.  These mechanisms are especially problematic for binaries
involving a supernova.  When a massive primary star evolves (prior to exploding), it
experiences significant mass loss.  Dynamical friction from the primary's wind
causes the secondary star to quickly in-spiral, thus conserving the angular momentum
of the system \citep[][]{vanbeveren98}.
	A minimum binary separation must be chosen to prevent mergers.
Unfortunately, the details of tidal dissipation and stellar merging are uncertain.
Thus we make only optimistic assumptions in our estimate of hyper-runaway ejection
rates.

\subsubsection{Hyper-Runaway Ejection Rate}

	In the context of our HVSs, a runaway must have a velocity exceeding 400 km
s$^{-1}$ to be confused with a HVS.  The orbital velocity of the secondary star in a
binary is
	\begin{equation} \label{eqn:vsec}
 v_{\rm sec} = \frac{2 \pi a}{(1+q)P},
	\end{equation} where $a$ is the separation of the two stars, $q=M_2/M_1$ is
the mass ratio, and $P$ is the orbital period.  We insert Kepler's third law $P^2 =
4 \pi^2 a^3 / (GM)$ into Equation \ref{eqn:vsec}, set $v_{\rm sec} \ge 400$ km
s$^{-1}$, and find that the progenitor binary system must have
	\begin{equation} \label{eqn:ma}
 M/a \ge 0.84 (1+q)^2 ~ ({\rm M}_{\sun}/{\rm R}_{\sun})
	\end{equation} to produce a $\ge$400 km s$^{-1}$ runaway.  Here, $M$ is the
total mass of the binary.  We will optimistically assume that the secondary fills
its Roche-lobe and has a radius of $\sim$0.4$a$.  Because a 3 M$_{\sun}$ star has a
3 R$_{\sun}$ radius, a binary must have $a\ge8$ R$_{\sun}$ and a primary star with
mass $\ge$10 M$_{\sun}$ to produce the requisite 3 M$_{\sun}$ runaway ejected at
$\ge$400 km s$^{-1}$ (Equation \ref{eqn:ma}).

	Known HVSs have travel times spanning 200 Myr (see Figure \ref{fig:travel}).  
The star formation rate in the solar neighborhood is 0.5 M$_{\sun}$ yr$^{-1}$
\citep{lada03}.  Thus $10^8$ M$_{\sun}$ of stars have formed in the disk in the past
200 Myr.

	A standard Salpeter initial mass function \citep{salpeter55}, integrated
from 0.1 M$_{\sun}$ to 100 M$_{\sun}$ and normalized to $10^8$ M$_{\sun}$, predicts
$5.4\times10^5$ stars with masses 10-100 M$_{\sun}$.
	All O and B stars are in binaries, and a third of the binaries are twins
\citep{kobulnicky07}.  Thus $1.8\times10^5$ of massive primaries are not twins.  
Assuming the secondaries have a Salpeter mass function \citep{kobulnicky07}, there
are $\sim$600 secondaries with mass 3-4 M$_{\sun}$.  Given a log-normal distribution
of binary separations, $\sim$5\% of binaries have 8-20 R$_{\sun}$ semi-major axes
enabling a $\ge$400 km s$^{-1}$ ejection.  Thus we expect $\sim$30 3-4 M$_{\sun}$
runaways ejected at $\ge$400 km s$^{-1}$ in the past 200 Myr.

	The Galactic disk has an exponential stellar density profile with radial
scale length 2.4 kpc \citep{siegel02}.  Thus the region of the outer disk $10<R<20$
kpc, despite containing most of the disk's area, contains no more than 10\% of the
disk's stars if we optimistically normalize the density over $5<R<20$ kpc.  Thus we
predict only $\sim$3 possible 3-4 M$_{\sun}$ hyper-runaways ejected from the outer
disk in the past 200 Myr.

	However, potential models of the Milky Way show that a star traveling with
400 km s$^{-1}$ in the region $10<R<20$ kpc is bound (see Figure \ref{fig:travel}).  
To achieve an unbound velocity, a runaway must be ejected at 400 km s$^{-1}$ in the
direction of Galactic rotation.  Assuming that runaways are ejected in random
directions, no more than 10\% of ejections will be in the direction of Galactic
rotation.  Thus we predict $\sim$0.3 hyper-runaways with mass 3-4 M$_{\sun}$ were
ejected in the past 200 Myr, preferentially found at low Galactic latitudes.

	The ejection rate from binary-binary encounters is even smaller, because it
depends on the joint probability of colliding two compact binaries.  A cluster of
$10^5$ stars following a Salpeter mass function contains 330 3-4 M$_{\odot}$ stars
and 190 10-100 M$_{\odot}$ stars.  We optimistically assume that the 330 3-4
M$_{\odot}$ stars are in 330 different binaries, and that the 10-100 M$_{\odot}$
stars are in 130 binaries \citep[because a third of them are
twins,][]{kobulnicky07}. A log-normal distribution of binary separation suggests
that $\sim$5\% of the binaries have 8-20 R$_{\sun}$ semi-major axes enabling a
$\ge$400 km s$^{-1}$ ejection.  This reduces the number of relevant binaries
containing 3-4 M$_{\odot}$ and 10-100 M$_{\odot}$ stars to 17 and 7, respectively.
If mass segregation puts all of the compact binaries in the central 0.1 pc, then the
space density of compact binaries containing a 10-100 M$_{\odot}$ star is
2$\times10^3$ pc$^{-3}$.
	If we assume a velocity dispersion of 1 km s$^{-1}$ and a cross-section for
collision of $r=20$ R$_{\odot}$, then the ejection rate of 3-4 M$_{\sun}$ $\ge$400
km s$^{-1}$ runaways is $\sim$$2\times10^{-8}$ Myr$^{-1}$ per cluster.  Because 10
M$_{\odot}$ stars live $\sim$20 Myr \citep[e.g.][]{schaller92}, and because we
require $\sim$3000 clusters to reach 10$^8$ M$_{\odot}$, we expect $\sim$10$^{-3}$
binary-binary runaways with mass 3-4 M$_{\sun}$ and ejection velocity $\ge$400 km
s$^{-1}$ in the past 200 Myr.  However, these binary-binary runaways are subject to
the same outer disk and Galactic rotation constraints as the supernova runaways.  
Thus we predict that only $\sim$10$^{-5}$ 3-4 M$_{\sun}$ {\it hyper}-runaways were
ejected by binary-binary encounters in the past 200 Myr.

	In contrast, our HVS discoveries imply $96\pm20$ unbound 3-4 M$_{\sun}$
stars were ejected over the same time period.  We conclude that 3-4 M$_{\odot}$ HVSs
ejected from the Galactic Center are $\gtrsim$100 times more common than
hyper-runaways of the same mass.

	Hyper-runaways are rare because of the rarity of massive stars and compact
binaries, and the requirement to avoid merging the compact binary progenitors.  
	Hyper-runaways are also preferentially located at low Galactic latitudes.  
While it is possible for a hyper-runaway to be confused with an HVS in the absence
of proper motions, the observed $\sim$3 M$_{\odot}$ unbound stars are almost
certainly HVSs ejected by the central MBH.

\begin{deluxetable*}{lccccccccl}           % TABLE OF KNOWN HVSs
\tabletypesize{\scriptsize}
\tablewidth{0pt}
\tablecaption{HVS SURVEY STARS WITH V$_{rf}>+275$ KM S$^{-1}$\label{tab:hvs}}
\tablecolumns{10}
\tablehead{
  \colhead{ID} & \colhead{Type} & \colhead{$M_V$} & \colhead{$V$} &
  \colhead{$R_{GC}$} & \colhead{$l$} & \colhead{$b$} & \colhead{$v_{\sun}$} &
  \colhead{$v_{rf}$} & \colhead{Catalog} \\
  \colhead{} & \colhead{} & \colhead{mag} & \colhead{mag} &
  \colhead{kpc} & \colhead{deg} & \colhead{deg} & 
  \colhead{km s$^{-1}$} & \colhead{km s$^{-1}$} & \colhead{}
}
	\startdata
\cutinhead{HVSs}
HVS1  &  B  & -0.3 & 19.83 & 111 & 227.33 & +31.33 & 840 & 696 & SDSS J090744.99+024506.9$^1$ \\
HVS2  & sdO & +2.6 & 19.05 &  26 & 175.99 & +47.05 & 708 & 717 & US 708$^2$                   \\
HVS3  &  B  & -2.7 & 16.20 &  62 & 263.04 & -40.91 & 723 & 548 & HE 0437-5439$^3$             \\
HVS4  &  B  & -0.9 & 18.50 &  82 & 194.76 & +42.56 & 611 & 566 & SDSS J091301.01+305119.8$^4$ \\
HVS5  &  B  & -0.3 & 17.70 &  45 & 146.23 & +38.70 & 553 & 649 & SDSS J091759.48+672238.3$^4$ \\
HVS6  &  B  & -0.3 & 19.11 &  78 & 243.12 & +59.56 & 626 & 528 & SDSS J110557.45+093439.5$^5$ \\
HVS7  &  B  & -1.1 & 17.80 &  60 & 263.83 & +57.95 & 529 & 416 & SDSS J113312.12+010824.9$^5$ \\
HVS8  &  B  & -0.3 & 18.09 &  53 & 211.70 & +46.33 & 489 & 407 & SDSS J094214.04+200322.1$^6$ \\
HVS9  &  B  & -0.3 & 18.76 &  68 & 244.63 & +44.38 & 628 & 485 & SDSS J102137.08$-$005234.8$^6$ \\
HVS10 &  B  & -0.3 & 19.36 &  87 & 249.93 & +75.72 & 478 & 432 & SDSS J120337.85+180250.4$^6$ \\
HVS11 &  A  & +0.6 & 19.70 &  70 & 238.76 & +40.63 & 482 & 336 & SDSS J095906.48+000853.4 \\
HVS12 & A/BHB & +0.6 & 19.76 &  70 & 247.11 & +52.46 & 552 & 429 & SDSS J105009.60+031550.7 \\
HVS13 &  B  & -0.3 & 20.16 & 125 & 251.65 & +50.64 & 575 & 443 & SDSS J105248.31$-$000133.9 \\
HVS14 &  B  & -0.3 & 19.89 & 112 & 241.78 & +53.20 & 532 & 416 & SDSS J104401.75+061139.0 \\
HVS15 &  B  & -0.3 & 19.33 &  85 & 266.51 & +55.92 & 463 & 343 & SDSS J113341.09$-$012114.2 \\
HVS16 &  B  & -0.3 & 19.49 &  90 & 285.86 & +67.38 & 443 & 367 & SDSS J122523.40+052233.8 \\
\cutinhead{Possible HVSs}
      &  A  & +1.3 & 20.18 &  65 & 162.98 & +46.34 & 235 & 279 & SDSS J094014.56+530901.7 \\
      &  A  & +0.6 & 19.95 &  79 & 155.49 & +49.45 & 228 & 289 & SDSS J101359.79+563111.7 \\
      &  A  & +0.6 & 19.54 &  59 &   0.14 & +69.28 & 279 & 289 & SDSS J140306.54+145005.0 \\
      &  A  & +0.6 & 20.30 &  83 &  39.36 & +50.87 & 193 & 293 & SDSS J154556.10+243708.9 \\
\cutinhead{Possible Bound HVSs}
      &  B  & -0.3 & 17.38 &  40 & 189.17 & -48.75 & 314 & 279 & SDSS J032054.69$-$060616.0 \\
      &  B  & -0.3 & 18.56 &  66 & 196.07 & +23.21 & 361 & 298 & SDSS J074950.24+243841.2 \\
      &  B  & -0.3 & 17.43 &  42 & 160.45 & +34.20 & 229 & 288 & SDSS J081828.07+570922.1 \\
      &  B  & -0.3 & 18.23 &  57 & 186.30 & +42.16 & 306 & 285 & SDSS J090710.08+365957.5 \\
      &  B  & -0.3 & 18.13 &  51 & 251.20 & +54.36 & 451 & 331 & SDSS J110224.37+025002.8 \\
      &  B  & -0.3 & 18.31 &  53 & 274.88 & +57.45 & 424 & 310 & SDSS J115245.91$-$021116.2 \\
      &  B  & -0.3 & 17.64 &  39 &  65.34 & +72.37 & 228 & 298 & SDSS J140432.38+352258.4 \\
      &  A  & +0.0 & 18.56 &  49 & 357.16 & +63.62 & 284 & 290 & SDSS J141723.34+101245.7 \\
	\enddata
\tablerefs{ (1) \citet{brown05}; (2) \citet{hirsch05}; (3) \citet{edelmann05};
(4) \citet{brown06}; (5) \citet{brown06b}; (6) \citet{brown07b} }
 \end{deluxetable*}

\section{NEW HYPERVELOCITY STARS}

\subsection{Six Unbound HVSs}

\begin{figure}		% FIGURE 4: HVS SPECTRUM
 \plotone{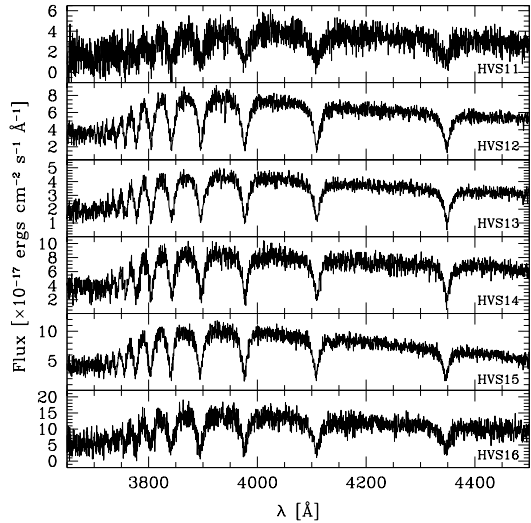}
 \caption{ \label{fig:spec}
	Observed MMT spectra of the 6 new HVSs.}
 \end{figure}

	Here we describe the 6 unbound HVSs newly discovered in our survey.  The
first two stars are of later spectral type than the HVS discoveries in our previous
targeted survey \citep{brown06, brown06b, brown07a, brown07b}.

	SDSS J095906.48+000853.40, hereafter HVS11, has an A1 spectral type, a
$+482\pm19$ km s$^{-1}$ heliocentric radial velocity, and a minimum velocity of
$+336$ km s$^{-1}$ in the Galactic rest frame.  An A-type spectral classification is
supported by a strong $\lambda3933$ Ca{\sc ii} K line in the spectrum (Figure
\ref{fig:spec}).  HVS11 is the reddest HVS identified to date, with a broadband
color $(g'-r')_0=-0.256\pm0.028$.  A solar metallicity 2.5 M$_{\sun}$ main sequence
star has $M_V$(2.5 M$_{\odot}$) $\simeq+0.6$ \citep{schaller92}.  This luminosity
places HVS11 at an approximate Galactocentric distance $R=70$ kpc.

	Known HVSs are typically separated by 10\arcdeg\ - 20\arcdeg\ from the
nearest Local Group dwarf galaxy, however HVS11 is only 3\fdg9 from the Sextans
dwarf.  Any physical association with Sextans is very unlikely.  Sextans is
$1320\pm40$ kpc distant \citep{dolphin03} and has a heliocentric velocity of
$224\pm2$ km s$^{-1}$ \citep{young00}.  Thus HVS11 is moving towards the dwarf
galaxy with a relative velocity of +260 km s$^{-1}$.

	SDSS J105009.60+031550.67, hereafter HVS12, has an A0 spectral type, a
$+552\pm11$ km s$^{-1}$ heliocentric radial velocity, and a minimum velocity of
$+429$ km s$^{-1}$ in the Galactic rest frame.  HVS12 has a strong $\lambda3933$
Ca{\sc ii} K line and a higher $S/N$ spectrum than HVS11 (Figure \ref{fig:spec}).  
We measure a $0.8\pm0.1$ equivalent width of Ca{\sc ii} K.  Combining Ca{\sc ii} K
with the star's broadband color $(g'-r')_0=-0.307\pm0.039$, equivalent to
$(B-V)_0=-0.05$ \citep{clewley05}, we estimate [Fe/H]$=-0.5\pm0.7$
\citep{wilhelm99a}.  HVS12 is therefore consistent with being a solar metallicity
2.5 M$_{\sun}$ main sequence star, placing its at an approximate Galactocentric
distance $R=70$ kpc.

	SDSS J105248.31$-$000133.94, hereafter HVS13, has a $+575\pm11$ km s$^{-1}$
heliocentric radial velocity and a minimum velocity of $+443$ km s$^{-1}$ in the
Galactic rest frame.  Although its $(g'-r')_0=-0.295\pm0.034$ is nearly identical to
HVS12, HVS13 is 0.23 mag bluer in $(u'-g')_0$ and has a B9 spectral type (Figure
\ref{fig:spec}) consistent with a 3 M$_{\sun}$ main sequence star (see also Figure
\ref{fig:ugr}), the same spectral type observed for most of the other HVSs
in our survey.  At $g=20.18\pm0.02$, however, HVS13 is the faintest HVS discovered
to date.  A solar metallicity 3 M$_{\sun}$ main sequence star has $M_V\simeq-0.3$
\citep{schaller92}, which places HVS13 at $R=125$ kpc.

	SDSS J104401.75+061139.03, hereafter HVS14, has a B9 spectral type, a
$532\pm13$ km s$^{-1}$ heliocentric radial velocity, and a minimum velocity of $416$
km s$^{-1}$ in the Galactic rest frame.  Its broadband colors are consistent with a
3 M$_{\sun}$ main sequence star (Figure \ref{fig:spec}).  At $g=19.72\pm0.02$,
HVS14, like HVS13, has a very large Galactocentric distant $R=110$ kpc.

	SDSS J113341.09$-$012114.25, hereafter HVS15, has a B9 spectral type, a
$463\pm11$ km s$^{-1}$ heliocentric radial velocity, and a minimum velocity of $343$
km s$^{-1}$ in the Galactic rest frame.  HVS15 is the bluest of the new HVSs with
$(g'-r')_0= -0.346\pm0.031$, consistent with a 3 M$_{\sun}$ main sequence star.  We
previously classified HVS15 as a possibly bound HVS \citep{brown07b}, but in light
of its probable $R=85$ kpc distance the star is almost certainly unbound.  HVS15
would be located at $R=37$ kpc if it were a hot BHB star, yet its minimum rest frame
velocity would still be in excess of the \citet{xue08} Galactic escape velocity
estimate (see Figure \ref{fig:travel}).

	SDSS J122523.40+052233.85, hereafter HVS16, has a B9 spectral type, a
$443\pm14$ km s$^{-1}$ heliocentric radial velocity, and a minimum velocity of $367$
km s$^{-1}$ in the Galactic rest frame.  Its broadband colors are consistent with a
3 M$_{\sun}$ main sequence star, which places HVS16 at an approximate Galactocentric
distance $R=90$ kpc.  Similar to HVS15, HVS16 would be located about $R=44$ kpc if
it were a hot BHB star, but its minimum rest frame velocity would still be in excess
of the \citet{xue08} Galactic escape velocity estimate.  We conclude that both HVS15
and HVS16 are very likely unbound.

\subsection{Four Possible HVSs}	

	There are four HVS candidates with minimum rest frame velocities, main
sequence star distances, and BHB star distances that fall between the
\citet{kenyon08} and \citet{xue08} Galactic escape velocity models.  In other words,
these stars are unambiguously bound in the \citet{kenyon08} model, and unambiguously
unbound in the \citet{xue08} model.

	The four possible HVSs are SDSS J094014.56+530901.74, SDSS
J101359.79+563111.66, SDSS J141342.50+442550.03, and SDSS J154556.10+243708.94.  
All four stars have minimum rest frame velocities around +290 km s$^{-1}$ and the
spectral types of early A-type stars.  These four possible HVSs are systematically
redder than the other HVSs, consistent with 2 - 2.5 M$_{\sun}$ stars (see Figure
\ref{fig:ugr}).  Their main-sequence star and BHB distances range 55 - 85 kpc (see
Figure \ref{fig:travel}).

	Although the four stars are very possibly unbound, an HVS origin is not a
unique explanation for their velocities.  As explained above, runaway ejections can
``contaminate'' the low-velocity end of HVSs.  It is, however, more likely that
unbound low mass A stars are ejected by a MBH.  Proper motions are needed to
distinguish these four stars as HVSs or runaways.

\subsection{HVS Table}

	Table \ref{tab:hvs} lists the 26 stars in our HVS survey with $v_{rf}>+275$
km s$^{-1}$, plus HVS2 and HVS3 for completeness.  Magnitudes and radial velocities
are observed quantities, whereas luminosities and distances are inferred from
spectra and colors.  Columns include HVS number, stellar type, absolute magnitude
$M_V$, apparent $V$ magnitude derived from SDSS photometry, Galactocentric distance
$R$, Galactic coordinates $(l,b)$, heliocentric radial velocity $v_{\sun}$, minimum
Galactic rest-frame velocity $v_{rf}$ (not a full space velocity), and catalog
identification.  We report the weighted average velocity measurements for each
object.  Thus the velocities in Table \ref{tab:hvs} may vary slightly from earlier
work.

	We do not report errors in Table \ref{tab:hvs} because formal uncertainties
are misleadingly small compared to the (unknown) systematic errors.  For example,
our radial velocities have 11-17 km s$^{-1}$ uncertainties, but we have no
constraint on the proper motion component of the rest frame velocity $v_{rf}$.  The
luminosity estimates are precise at the 10\% level for main sequence stars.  
However, the luminosity estimates could be over-estimated by an order of magnitude
for post-main sequence stars.

\section{THE NATURE OF HYPERVELOCITY STARS}

	We argued previously that HVSs must be short-lived main sequence stars
\citep{brown07b}.  Follow-up observations have, remarkably, confirmed that four
B-type HVSs are main sequence stars:  HVS1 is a slowly pulsating B variable
\citep{fuentes06}, HVS3 is a 9 M$_{\sun}$ B star \citep{bonanos08, przybilla08},
HVS7 is a 3.7 M$_{\sun}$ Bp star \citep{przybilla08b}, and HVS8 is a rapidly
rotating B star \citep{lopezmorales08}.

	The identification of HVSs as main sequence stars is in stark contrast to
the halo stars in our survey, which are, presumably, evolved 0.6-1 M$_{\sun}$ stars
on the BHB.  BHB stars among the HVSs would be exciting, however,
because unbound BHB stars would allow us to probe the low-mass regime of HVSs.

	\citet{kenyon08} calculate the observable spatial and velocity distribution
of HVSs as a function of mass, and predict that BHB stars are a factor of $\sim$10
times less abundant than main sequence HVSs.  Roughly speaking, solar mass HVSs are
10 times more abundant than 3-4 M$_{\sun}$ HVSs, but spend 1\% of their lifetime in
the BHB phase with luminosity comparable to an 2.5 M$_{\odot}$ main sequence star.
Not all of our HVSs have the colors of a 2.5 M$_{\odot}$ main sequence star (see
Figure \ref{fig:ugr}).  Yet, given the 16 - 20 HVSs identified to date, the
predictions of \citet{kenyon08} imply there should be 1$\pm$1 BHB stars among our
HVSs.

\begin{figure}		% FIGURE 5:
 \plotone{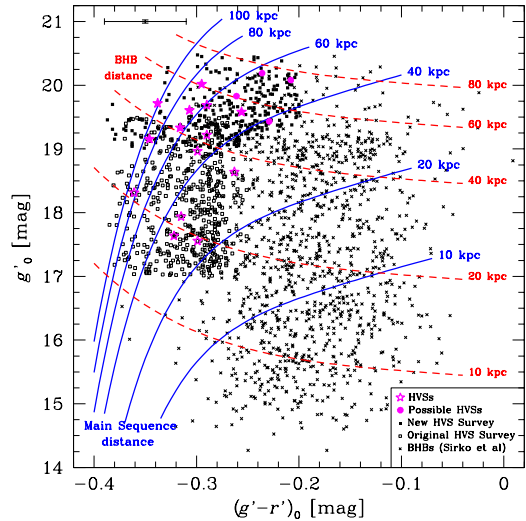}
 \caption{ \label{fig:colmags}
	Color-magnitude diagram showing that the original HVS survey stars ({\it
open squares}) are, by design, bluer than \citet{sirko04a} BHB stars ({\it
crosses}), while the new HVS survey stars ({\it solid squares}) are systematically
fainter than the BHB stars.  We plot SDSS DR6 uber-calibrated PSF magnitudes,
corrected for Galactic extinction, for all stars.  Lines indicate the distance for
main sequence stars with solar abundance \citep[solid lines,][]{girardi02,
girardi04} and for halo BHB stars with [Fe/H]$=-1.7$ \citep[dashed
lines,][]{brown08b}. }
 \end{figure}

\subsection{A Possible BHB HVS}

	BHB stars and main sequence stars are distinguished by surface gravity: low
surface gravity BHB stars have narrower Balmer lines (at a given effective
temperature) than high surface gravity main sequence stars.  Spectroscopic measures
of surface gravity work for temperatures cooler (redder) than $(\bv)_0=0$
\citep[e.g.][]{kinman94, wilhelm99a, clewley02}, colors that we probe now with our
new HVS survey.

	Applying the \citet{clewley02, clewley04} surface gravity measures to the
new HVSs, HVS12 is a possible BHB star.  The line width-shape technique indicates
HVS12 has low surface gravity, although the $D_{0.15}$-color technique is ambiguous
because of HVS12's blue $(\bv)_0=-0.05\pm0.04$ color.  If HVS12 is a BHB star with
[Fe/H]$=-1.5$, then it has $M_V$(BHB)$\simeq+1.2$ \citep{brown08b} and is located at
$R\simeq50$ kpc.  HVS12 is clearly unbound at this distance (see Figure
\ref{fig:travel}).

	Interestingly, HVS12 was previously classified as a BHB star by
\citet{sirko04a}.  \citet{sirko04a} published a sample of 1170 BHB stars observed as
mis-identified quasars and filler objects in the SDSS spectroscopic survey.  For
HVS12 they report a $532\pm35$ km s$^{-1}$ heliocentric velocity and a 0.847 Ca{\sc
ii} K equivalent width, consistent at the 1-$\sigma$ level with our own
measurements.  \citet{xue08} have recently re-analyzed the SDSS BHB sample and
combined it with additional targeted observations from the SEGUE survey.  
\citet{xue08} classify HVS12 as a possible BHB star, but exclude it from the their
rigorously selected BHB sample. High $S/N$ spectroscopy and photometry is required
to confirm the nature of HVS12.

	Although HVS12 is a clear outlier in the velocity distribution of BHB stars
\citep[see Figure 4 of][]{sirko04b}, HVS12 was not previously recognized as an
unbound star.  The median depth of the \citet{sirko04a} sample is $g'_0=17.35$,
corresponding to a distance of $\sim$20 kpc for a BHB star (see Figure
\ref{fig:colmags}).  A star with $v_{rf}\sim400$ km s$^{-1}$ at the median depth of
the \citet{sirko04a} sample is thus bound.  What appears to have escaped notice,
however, is that HVS12 is fainter, and bluer, than 96\% of the \citet{sirko04a} BHB
sample.

	Figure \ref{fig:colmags} compares the distribution of stars in our HVS
surveys and the \citet{sirko04a} BHB sample in a color-magnitude diagram.
	For reference, we draw lines of constant distance for main sequence stars
with solar abundance \citep[solid lines,][]{girardi02, girardi04} and for halo BHB
stars with [Fe/H]$=-1.7$ \citep[dashed lines,][]{brown08b}.
	Note that the color-magnitude selection of the new HVS survey is such that
every star has $R\gtrsim$40 kpc, whether a main sequence star or a BHB star (Figure
\ref{fig:colmags}).

	The presence of only one HVS among the 1170 \citet{sirko04a} BHB stars and
the 10224 \citet{xue08} BHB candidates shows the immense dilution due to stars 
in the Galactic halo.
	It is important to minimize contamination from foreground stellar
populations when looking for HVSs.
	Our surveys find HVSs because we target stars that are bluer and/or fainter
than the bulk of halo BHB stars.

\section{CONCLUSIONS}

	We describe a new targeted HVS survey, a spectroscopic survey of faint stars
$19<g'_0<20.5$ with early A-type and late B-type colors.  
	Recent observations confirm that 3 of our B-type HVSs are 3-4 M$_{\sun}$
main sequence stars.

	The observational signature of a HVS is its unbound velocity, which we
determine by comparing observed radial velocities and distances to Galactic
potential models.  We argue that the known properties of binaries and the rarity of
massive stars make hyper-runaways like HD 271791 rare.  A MBH ejection remains the
most plausible origin of unbound low-mass stars.

	Our HVS survey is 59\% complete and, combined with our original HVS survey,
shows a remarkable velocity distribution:  26 stars with $v_{rf}>+275$ km s$^{-1}$
and only 2 stars with $v_{rf}<-275$ km s$^{-1}$.  
	Here we report the discovery of 6 new unbound HVSs in excess of the
conservative escape velocity model of \citet{kenyon08}, and 4 additional unbound
HVSs in excess of the escape velocity model of \citet{xue08}.

	One of the new HVSs may be an evolved BHB star.  The \citet{kenyon08}
ejection models predict BHB HVSs are $\sim$10 times less abundant than the main
sequence HVSs in our survey, consistent with the existence of $1\pm1$ BHB stars in
our HVS sample.  Of course, the exact number of BHB HVSs depends on the mass
function of stars near the central MBH.
	BHB HVSs therefore have the potential to probe the low-mass regime of HVSs
and constrain the mass function of stars in the Galactic center.

	HVSs are fascinating because their properties are tied to the nature and
environment of the MBH that ejects them \citep{levin06, baumgardt06, merritt06,
ginsburg06, ginsburg07, demarque07, gualandris07, sesana06, sesana07, sesana07b,
sesana07c, lu07, kollmeier07, hansen07, perets07, perets07c, perets08b, perets08a,
sherwin08, svensson08, oleary08, lockmann08}, and their trajectories probe the dark
matter halo through which they move \citep{gnedin05, yu07, wu08, kenyon08}.  
	The angular distribution of HVSs on the sky reveals significant anisotropy
that may also be related to the Galactic potential \citep{brown08d}.
	Our ultimate goal is to find a statistical sample of $\sim$100 HVSs to
measure the distribution of HVS properties and discriminate HVS ejection models.

	~

\acknowledgements

	We thank M.\ Alegria and A.\ Milone for their assistance with observations
obtained at the MMT Observatory, a joint facility of the Smithsonian Institution and
the University of Arizona.  This project makes use of data products from the Sloan
Digital Sky Survey, which is managed by the Astrophysical Research Consortium for
the Participating Institutions.  This research makes use of NASA's Astrophysics Data
System Bibliographic Services.  This work was supported by the Smithsonian Institution.

{\it Facilities:} {MMT (Blue Channel Spectrograph)}

\appendix
\section{DATA TABLE}

	Table \ref{tab:qso} presents the 19 $z\sim2.4$ quasars and 9 DA white dwarfs
in our survey.  Table columns include RA and Dec coordinates (J2000), $g'$ apparent
magnitude, $(u'-g')_0$ and $(g'-r')_0$ color, and spectroscopic identification.

\begin{deluxetable}{rrcccc}           % TABLE OF QSOs and DA WDs
\tabletypesize{\small}
\tablewidth{0pt}
\tablecaption{QUASARS AND WHITE DWARFS\label{tab:qso}}
\tablecolumns{6}
\tablehead{
  \colhead{RA} & \colhead{Dec} & \colhead{$g'$} & \colhead{$(u'-g')_0$} &
  \colhead{$(g'-r')_0$} & \colhead{Type} \\
  \colhead{hrs} & \colhead{deg} & \colhead{mag} & \colhead{mag} &
  \colhead{mag} & \colhead{}
}
	\startdata
 4:06:24.10 & -4:19:34.0 &  20.594 &  0.763 & -0.268 &  QSO \\
 8:05:30.10 &  2:18:45.0 &  20.448 &  0.804 & -0.225 &  QSO \\
 8:06:21.42 & 33:38:32.8 &  19.993 &  0.649 & -0.244 &  WD  \\
 8:23:36.89 &  1:52:55.9 &  20.445 &  0.873 & -0.220 &  QSO \\
 9:03:21.90 & 49:51:49.0 &  20.315 &  0.639 & -0.276 &  WD  \\
 9:19:14.83 & 12:52:06.0 &  20.014 &  0.725 & -0.250 &  WD  \\
 9:22:11.31 & 45:57:19.4 &  20.244 &  0.900 & -0.270 &  QSO \\
 9:52:18.52 & 33:24:46.4 &  20.110 &  0.638 & -0.250 &  WD  \\
10:00:52.77 & 40:51:23.3 &  20.078 &  1.021 & -0.216 &  QSO \\
10:18:11.91 & 50:16:00.9 &  20.422 &  0.611 & -0.292 &  QSO \\
10:42:58.03 & 28:30:33.4 &  20.361 &  1.000 & -0.344 &  QSO \\
10:45:01.96 & -1:19:46.7 &  20.284 &  0.673 & -0.263 &  QSO \\
11:02:16.15 &  7:54:20.7 &  19.773 &  0.853 & -0.211 &  QSO \\
11:10:19.82 & 59:14:59.3 &  19.753 &  0.887 & -0.291 &  QSO \\
11:18:54.44 & 30:27:09.9 &  20.366 &  0.757 & -0.239 &  QSO \\
11:41:02.74 & 42:20:34.0 &  20.561 &  0.847 & -0.337 &  QSO \\
11:56:56.37 & 22:41:55.3 &  20.141 &  0.794 & -0.271 &  QSO \\
11:57:26.83 & -1:29:14.9 &  19.872 &  0.610 & -0.271 &  WD  \\
12:15:53.64 & 34:23:18.2 &  20.058 &  0.831 & -0.223 &  QSO \\
12:29:08.32 & 49:58:27.2 &  20.416 &  0.961 & -0.213 &  QSO \\
13:07:54.94 & 48:35:25.9 &  20.168 &  0.661 & -0.238 &  QSO \\
14:13:42.51 & 44:25:50.0 &  20.457 &  0.807 & -0.224 &  QSO \\
14:22:00.74 & 43:52:53.2 &  19.821 &  0.715 & -0.271 &  WD  \\
15:33:00.04 & 49:29:48.3 &  19.278 &  0.611 & -0.248 &  WD  \\
15:43:24.56 & 36:26:49.5 &  19.599 &  0.926 & -0.275 &  QSO \\
15:58:51.85 & 22:21:59.9 &  19.671 &  0.817 & -0.292 &  QSO \\
16:11:27.35 & 26:56:10.9 &  19.977 &  0.647 & -0.305 &  WD  \\
17:40:43.32 & 67:24:41.5 &  19.680 &  0.663 & -0.268 &  WD  \\
	\enddata
 \end{deluxetable}

	% REFERENCES 
\clearpage

\end{document}